\documentclass[onecollarge,referee]{svjour2}

\usepackage{graphicx}
\usepackage{bm}
\usepackage{amsfonts}
\usepackage{amssymb,amsmath}
\usepackage[justification=centering]{caption}
\usepackage[bookmarks=true,colorlinks=true,linktocpage=true,backref=true]{hyperref}

\begin{document}


\title{An Inhomogeneous Space-Time Patching Model\\ 
Based on a Nonlocal and Nonlinear Schr\"odinger\\ 
Equation}

\titlerunning{An Inhomogeneous Space-Time Patching Model }

\author{Christine C. Dantas}

 \institute{Divis\~ao de Materiais (AMR), Instituto de Aeron\'autica e Espa\c co (IAE), Departamento de Ci\^encia e Tecnologia Aeroespacial (DCTA), Brazil, \email{christineccd@iae.cta.br}}

\journalname{Foundations of Physics}

\date{Submitted: \today}

\maketitle

\begin{abstract}
We consider an integrable, nonlocal and nonlinear, Schr\"odinger equation (NNSE) as a model for building space-time patchings in inhomogeneous loop quantum cosmology (LQC).  We briefly review exact solutions of the NNSE, specially those obtained through ``geometric equivalence'' methods. Furthemore, we argue that the integrability of the NNSE could be linked to consistency conditions derived from LQC, under the assumption that the patchwork dynamics behaves as an integrable many-body system.

\end{abstract}




\normalsize


\section{Introduction}

Quantum cosmology is the application of quantum gravity to cosmological problems, particularly to the initial conditions of the Universe. Unfortunately, a complete theory of quantum gravity has not been found yet,  mainly due to the lack of experimental verification and guidance. Currenlty, the most developed approach in the canonical quatization framework is loop quantum gravity (LQG) \cite{LQG}, wherein loop quantum cosmology (LQC) \cite{LQC} \cite{BOUNCE} is one of its most explored applications.

There are currently many candidate theories of quantum gravity, and related conceptual and technical issues are evidently transfered to quantum cosmology, which also presents its own internal problems. However, it is believed that, in its development, quantum cosmology may provide insights about the nature of space-time, novel mathematical techniques, as well as the possibility of testing potential quantum gravity signatures observationally \cite{SIG}.

Very recently, the traditional minisuperspace quantization used in LQC has been subject to a conceptual re-evaluation, and the view of quantizing a single-patch space and combining patches in a dynamical ``many-patch system'' has been emphasized as a more fundamental and relevant alternative  \cite{Boj15}. Such a view attempts to connect the LQC formalism with broadly available and tested condensed matter techniques, with corresponding adaptations. In particular, it is expected that the many-patch dynamics is effectively described by a nonlinear single-patch dynamics of states, wherein each single patch is quantized according to the quantum gravity theory considered. Techniques for treating nonlinearity will be necessary in these developments. 

Methods for obtaining solutions to certain classes of nonlinear problems in various fields have been developed in the last decades, as for instance the inverse scattering transform \cite{IST}. On the other hand, deep mathematical connections between {\sl geometry} and certain integrable partial differential equations are known to exist, and have been discussed in a voluminous literature (e.g., Ref. \cite{GEO} and references therein). Not only explicit solutions to can be found, but also insights on the nature of integrability can be obtained (e.g., Ref. \cite{INT}).  

In particular, the well-known and dynamically rich (1+1)-dimensional Heisenberg ferromagnetic spin systems \cite{Lak11} correspond to the hierarchy of {\sl nonlinear Schr\"odinger equations} (NSE) \cite{Lak7677}, through the so-called  {\sl geometric equivalence} (GE) procedures (e.g. Ref. \cite{GE-REFS}). The GE method uses embedding constructions in $\mathbb{E}^3$ involving rigid moving curves (the origins and the vast scope of the method can be appreciated, e.g., Ref. \cite{GEO}). Discrete Heisenberg ferromagnetic spin systems and their integrability have also been subject of study in that context \cite{DISC}.  Higher order and higher-dimensional spin systems have also been investigated, including  inhomogeneous spin systems \cite{NLS-GEN}, with or without nonlocal terms, the former referred as the {\sl nonlocal NSE}, (hereon, NNSE; e.g., Refs. \cite{NONLOC}, \cite{Bal85}).  Generalizations of the rigid moving curves method by surface theory considerations in more general spaces have also been developed (e.g., Refs. \cite{Cie93}, \cite{Bal96}, \cite{Bal97}, \cite{SURF}).

Our preliminary considerations \cite{Dan13}, connecting Heisenberg spin systems with the qualitative expectations of LQC discrete evolution equations, addressed similarities between the latter and the NSE. A framework addressing nonlinear LQC, considering condensed matter ideas for space-time patchings, was further proposed in Ref. \cite{Boj12}, where the need of a nonlocal model was pointed out. Independent developments with analogous results, using group field theory condensates, have also been recently investigated (e.g. Ref. \cite{SIM}). 

The present work continues those studies, now examining the NNSE as a model for the dynamics of inhomogeneous space-time patchworks. 
Briefly, we propose an alternative formulation for the single-patch wave function space-time dynamics (c.f. Eqs. \ref{NNSE}, \ref{DNNSE}, and Sec. \ref{Sec_Motiv}), considering our previous investigations (c.f.,  Refs. \cite{Dan13}, \cite{Boj12}). Specifically, this proposal continues and generalises the developments outlined in Ref. \cite{Dan13} and is motivated by the proposal given in Ref. \cite{Boj12}. Sec. \ref{Sub_Compare} discusses in what way the present formulation extends/corresponds to the latter proposal.  Furthermore, it presents a methodological procedure leading to exact solutions, representing a useful tool for a systematic investigation of various scenarios in quantum cosmology. This methodology, based on geometric concepts, may be explored in a systematic study of the solution surfaces and their distortions, relating them to consistency conditions imposed by quantum gravity corrections. This may contribute to the advancement of current conceptual issues in quantum cosmology, specially through identifying constraints for potential observables in quantum cosmology.

This paper is organized as follows: in Section \ref{Sec_NNSE} we overview the generalised (nonlocal, inhomogeneous) integro-differential nonlinear Schr\"odinger equation (NNSE) and some classes of known exact solutions, including solitons.  In Sec. \ref{Sec_Motiv}, we present the motivations and interpretation of the NNSE in the context of a inhomogeneous LQC model. In Sec. \ref{Sec_GETool}, we present a brief note on the integrability of the NNSE in relation to certain consistency conditions arising in LQC. We conclude our paper in Sec. \ref{Sec_Conc}. The Appendices outline the main ideas of the GE method through the closely related formalisms of ``moving curves'' (App. \ref{App_SCF}) and surface theory (App. \ref{App_surface_theory}).


\section{Overview of the Nonlocal and Nonlinear Schr\"odinger Equation \label{Sec_NNSE}}

In this section, we review a generalized (nonlocal, inhomogeneous) form of the nonlinear Schr\"odinger equation (NSE), denoted here by the nonlocal and nonlinear Schr\"odinger equation (NNSE; c.f. Refs. \cite{NLS-GEN},\cite{Bal85},\cite{Cie93}, \cite{Bal96}, \cite{Bal97}). We also mention the corresponding discrete evolution equations of these equations (denoted by DNSE and DNNSE, respectively; c.f. Ref. \cite{DISC}). Relatively simple geometric equivalence (GE) procedures show that Heisenberg ferromagnetic spin systems correspond to the NSE and related forms \cite{Lak7677}, \cite{GE-REFS}.  Appendix \ref{App} gives an overview of these geometric procedures; in particular a proof of the equivalence between the NSE and the simplest Heinsenberg spin chain is reviewed in the Appendix \ref{App_SCF}.

\subsection{Some Forms of Nonlinear Schr\"odinger Equations \label{Sub_Forms}}

The discrete nonlinear Schr\"odinger equation (DNSE) \cite{DISC} is a nonlinear differential-difference equation for the wave function $q_n$, $n \in \mathbb{Z}$,
\begin{equation}
i {\partial \over \partial t}q_n = (q_{n+1} -  2q_n + q_{n-1} ) \pm |q_n|^2 (q_{n+1} + q_{n-1} ) , ~~~~~~{\rm (DNSE)}\label{DNSE}
\end{equation}

\noindent where the second term on the right hand side represents the nonlinearity. We have adopted the discretization used by Ablowitz et al. (2004, in Ref. \cite{Lak7677});  Hoffmann (2000, in Ref. \cite{DISC}) argues that the latter form is a more natural choice from a geometrical point of view (see other discretizations of the NSE in that paper and references therein).

The continuous limit of the DNSE, obtained when the
spin and coupling vary slowly over the lattice distance, is given by the nonlinear Schr\"odinger equation (NSE) (c.f. Ablowitz et al. in \cite{Lak7677}):
 
\begin{equation}
i {\partial \over \partial t}q(x,t) + {\partial^2 \over \partial x^2}q(x,t)  + 2 |q(x,t)|^2 q(x,t) = 0. ~~~~~~{\rm (NSE)} \label{NSE}
\end{equation} 

\noindent Note that the nonlinearities of the NSE and DNSE are localized. 

A generalized (nonlocal, inhomogeneous) integro-differential nonlinear Schr\"odinger equation (NNSE)  \cite{NLS-GEN},\cite{Bal85},\cite{Cie93}, \cite{Bal96}, \cite{Bal97} can be written as:
\begin{equation}
i {\partial \over \partial t} q(x,t) + {\partial^2 \over \partial x^2}[f(x) q(x,t)] + 2 q(x,t) \left (f(x)|q(x,t)|^2 + \int_{-\infty}^x {\partial f \over \partial y} |q(y,t)|^2 dy \right ) = 0,  ~~~~~~{\rm (NNSE)} \label{NNSE}
\end{equation}
\noindent where $f(x)$ is an inhomogeneity (coupling) $\mathcal{C}^2$ function. The inhomogeneity function may also vary in time\footnote{Certain self-consistent solutions in this case seem to imply unbounded and/or negative inhomogeneity functions, which may be not physically admissible in some cases (c.f. Ref. \cite{Cie93}).}. Note that nonlocality (the integral expression) is incorporated in the nonlinear term of the NNSE. This equation is equivalent to the inhomogeneous spin system in the continuum limit given by Eq. \ref{S_cont}. 

The NNSE represents the continuous limit of a related differential-difference (discrete), nonlocal and nonlinear equation (hereon, the DNNSE; compare it with the DNSE, Eq. \ref{DNSE}):
\begin{equation}
i {\partial \over \partial t}q_n =  (fq)_{n+1} - 2 (fq)_n + (fq)_{n-1} 
\pm    \left  ( f_n | q_n | ^2  + \sum_m \Delta f_{n,m} | q_m| ^2 \right ) (q_{n+1} + q_{n-1}), ~~~~~~{\rm (DNNSE)}\label{DNNSE}
\end{equation}
\noindent where $\Delta f_{n,m}$ is a small-amplitude variation of the inhomogeneity function over some lattice neighborhood. 

\subsection{Soliton Solutions of the NNSE \label{Sub_Soliton}}

A class of soliton solutions to the NNSE (Eq. \ref{NNSE}) can be  obtained  by the use of well-known inverse scattering transform (IST) techniques \cite{IST}, \cite{Bal85}.  Other exact solutions \cite{Bal96}, \cite{Bal97} can be found through surface theory (c.f. Appendix \ref{App_surface_theory}), to be addressed in the next subsection. 

\begin{center}
\begin{figure}[htbp]
\centering
\includegraphics[width=0.7\columnwidth]{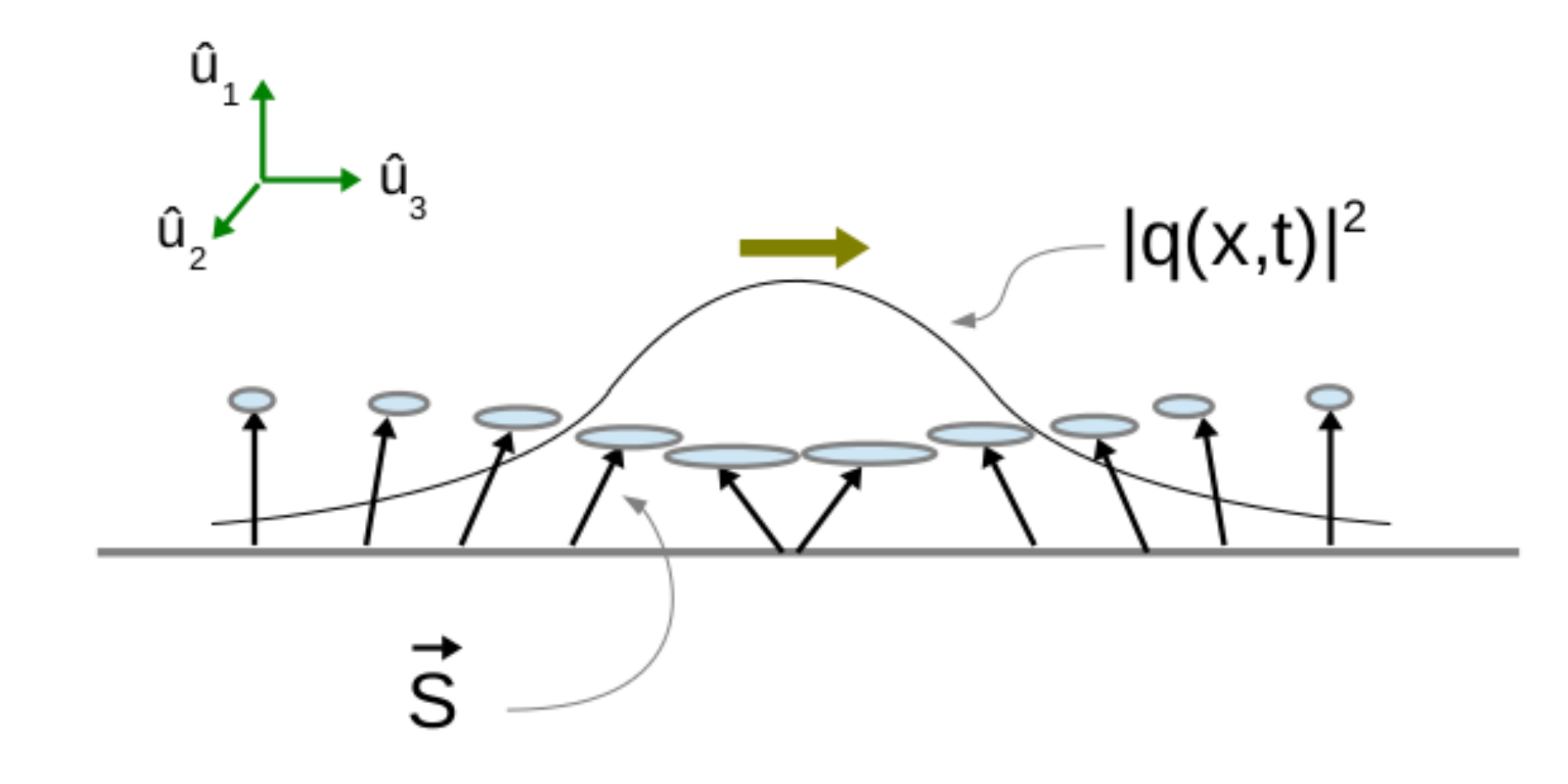}
\caption{\label{Fig_soliton} Soliton solution superposed on a Heisenberg spin chain (adapted from Murugesh \& Lakshmanan, 2005 \cite{GE-REFS}).}
\end{figure}
\end{center}

Fig. \ref{Fig_soliton} illustrates a propagating soliton, given by the profile $|q(x,t)|^2$, representing a class of exact solutions of the (homogeneous) NSE (Eq. \ref{NSE}). Also shown in that figure is the corresponding (1+1)-dimensional Heisenberg spin chain, where each spin vector rotates about the vertical axis, say, $\vec{\hat{u}}_1$. The soliton profile then measures the spin component projected onto the $\vec{\hat{u}}_2$--$\vec{\hat{u}}_3$ plane.

In the case of the NNSE (Eq. \ref{NNSE}) the soliton will propagate in an inhomogeneous media. An interesting soliton solution to the NNSE was found by Balakrishnan \cite{Bal85}:

\begin{equation}
q(x,t) = g(x) Q\left [ y(x),t \right ] , \label{Soliton_Sol}
\end{equation}
\noindent with  
\begin{equation}
y(x) \equiv \int_0^x g(x^{\prime}) dx^{\prime} \equiv \int_0^x {1 \over f(x^{\prime})^{1 \over 2}} dx^{\prime}, \label{Soliton_y}
\end{equation}
where we have choosen a simplified form ($\mu =0$ and $\mu_0=1$ in Eq. 30 of Ref. \cite{Bal85}), for the present exposition. The inhomogeneous function is specified by $f(x)$ above. The functional $Q$ is given by:
\begin{equation}
Q\left [ y(x),t \right ] = i c^* {\rm sech} \left \{ \theta [y(x),t] \right \} \exp \left \{-i \Phi [y (x),t ] \right \}, \label{Soliton_Q}
\end{equation}
\noindent with:
\begin{equation}
\theta\left [ y(x),t \right ]  = - 2 y(x) \eta - 16 \mu_0 \eta \xi t + \ln \left ({|c|\over 2 \eta} \right ), \label{Soliton_theta}
\end{equation}
\noindent and
\begin{equation}
\Phi\left [ y(x),t \right ]  = 2y(x)\xi - 8 \mu_0 \left (\xi^2 - \eta^2 \right ) t- i \ln \left ({|c|\over 2 \eta} \right ). \label{Soliton_phi}
\end{equation}
\noindent In the equations above, $\xi$ and $\eta$ are constants, given by the initial complex eigenvalue of the IST problem:
\begin{equation}
h_1(0) = \xi + i \eta, ~~~~ \eta > 0,
\end{equation}
\noindent and $c$ is a complex parameter specified by $h_1(0)$ (see Ref. \cite{Bal85}).

Localized (soliton) forms for some types of inhomogeneity (or coupling) functions can be found. For instance, solitons arise in the case of a  sech$^2$-shaped form,
\begin{equation}
f(x) =  {\rm sech}^2(\alpha x + \beta) , \label{Eq_fx}
\end{equation}
\noindent where $\alpha$ and $\beta$ are arbitrary constants. For a simplified expression ($\alpha =1$, $\beta = 0$), Eq. \ref{Soliton_Sol} then takes the functional form:
\begin{equation}
q(x,t) \sim  \cosh(x) {\rm sech} \left \{ \theta [\sinh(x),t] \right \} \exp \left \{-i \Phi [\sinh (x),t ] \right \} ~~~~~~\in \mathbb{C}. \label{Soliton_Q_fx}
\end{equation}
\noindent It represents a soliton with an envelope shape ($|q(x,t)|^2$) which is similar to that of the inhomogeneity distribution (Eq. \ref{Eq_fx}). This solution is interesting because it represents a lossless propagation in the medium (the energy density of the spin system associated with the soliton also has a sech$^2$ form). The inhomogeneity plays the role of a potential barrier, where a ``nonlinear tunneling'' may occur \cite{Bal85}.

\subsection{Exact solutions of the NNSE from surface theory \label{Sec_GE_NNSE}}

One of the first works addressing the NSE in the context of surface theory (Sym \& Wesselius 1987, in Ref. \cite{NLS-GEN}) established a method to find self-consistent solutions starting from given parametric surfaces of rotation in $\mathbb{E}^3$, which was subsequently extended by Cie\'sli\'nski et al. \cite{Cie93} (various explicit solutions can be found in this paper). A few years later, through a different approach, Balakrishnan \& Guha \cite{Bal96} (see also Ref. \cite{Bal97}) applied surface theory to the inhomogeneous spin system and generalized the previous results. Surface theory is reviewed in the Appendix \ref{App_surface_theory}, following the derivation of the latter authors (for the Hasimoto transformation in Theorem \ref{T1} below, also check Eq. \ref{hasimoto_transf}).

We summarize the main results found by Balakrishnan \& Guha in the form of the following theorems:

\begin{theorem} \label{T1}
The Hasimoto transformation,
\begin{equation}
q(x,t) = {1 \over 2} L \exp \left ( i \int MG^{-1/2} dx \right ), \label{has}
\end{equation}
\noindent is an exact solution of the NNSE (Eq. \ref{NNSE}), where $G$, $L$ and $M$ are surface metric functions that obey the Gauss-Mainardi-Codazzi equations (c.f. Eqs. \ref{Gauss_Eq}--\ref{MC2_Eq}), with an additional constraint imposed by the inhomogeneity function (c.f. Eq. \ref{f_constraint_Eq}).
\end{theorem}

\begin{theorem} \label{T2}
Let $\Phi \equiv \Phi(x)$ be an arbitrary $\mathcal{C}^2$ function. Then an exact, time-independent  solution to the NNSE (Eq. \ref{NNSE}), $q(x,t) = q(x)$, is given by Eq. \ref{has}, with time-independent surface parameters obeying:
\begin{equation}
L = - \Phi /f , \label{L}
\end{equation}
\begin{equation}
M = C_0 \Phi^{-1}, \label{M}
\end{equation}
\begin{equation}
N = f (\Phi_{xx} - C_0 \Phi^{-3}), \label{N}
\end{equation}
\begin{equation}
G = \Phi^2 ~,  \label{G}
\end{equation}
\noindent where the inhomogeneous function is time-independent, $f \equiv f(x)$, and satisfies:
\begin{equation}
f(x) = \pm \Phi (2A_0 -\Phi_x^2 -C_0 \Phi^{-2})^{1 \over 2}/ (C_0 \Phi^{-3} - \Phi_{xx}), \label{f_eq}
\end{equation}
\noindent with the restrictions
\begin{equation}
2A_0 -\Phi_x^2 -C_0 \Phi^{-2}>0,
\end{equation}
\begin{equation}
C_0 \Phi^{-3} - \Phi_{xx} \neq 0.
\end{equation}
\end{theorem}

Therefore, for {\it time-independent metrics}, Theorem \ref{T2} states that if the coupling/inhomogeneous function $f(x)$ can be written in terms of some function $\Phi(x)$, and its derivatives, then an exact solution to the NNSE is guaranteed to exist. Alternatively, one may start with some {\it arbitrary} function $\Phi(x)$, and Theorem \ref{T2} gives the inhomogeneous function and corresponding surfaces parameters that lead to a solution to the NNSE. This is a remarkable result. 

The following cases are highlighted when Theorem \ref{T2} holds (time-independent metrics):

1) $M=0$ (and assuming $C_0=0$; c.f. Eq \ref{M}): 

\hspace{1cm} --   This limit corresponds to the ``spins on meridians'' solutions of Cie\'sli\'nski et al. \cite{Cie93}. Such surfaces select purely real wave function solutions (as can be directly checked from the Hasimoto transformation, Eq. \ref{has}). Furthermore, the extrinsic curvature functions (Eqs. \ref{L}--- \ref{N}) correspond to the following surface of revolution, with $\Phi$ playing the role of the generator of revolution\footnote{Therefore, the explicit solution for the equivalent spin system, Eq. \ref{S_cont}, can be written as:
\begin{equation}
\vec{S}(x,t)  = \vec{r}_x = \left [ \left ({1 \over 2}A_0 -\Phi_x^2 \right ) ^{1 \over 2}~,~\Phi_x \cos t~,~\Phi_x \sin t \right ]. \label{spin_sol_eq}
\end{equation}}:
\begin{equation}
\vec{r}(x,t) = \left [ \int  ({1 \over 2}A_0 -\Phi_x^2)^{1 \over 2} dx~,~ \Phi \cos t~,~ \Phi \sin t \right ]. \label{surfrev_eq}
\end{equation}

2) $M \ne 0$ (which implies $C_0\ne 0$; c.f. Eq \ref{M}): 

\hspace{1cm} --  Explicit construction of $\vec{r}(x,t)$ is nontrivial and, to our best knowledge, an open problem. \\

Concerning {\it time-dependent metrics}, we summarize the following results. Cie\'sli\'nski et al. \cite{Cie93} have found, from a different point of view, some explicit solutions ``away from meridians'', that is, in terms of elementary functions, which includes time-dependent metrics. However, those time-dependent solutions are very complicated expressions and were either related to negative or unbounded coupling functions $f(x)$. Hence, the physical relevance of these particular models by Cie\'sli\'nski et al. is uncertain. On the other hand,  under the assumptions of Theorem \ref{T1}, time-dependent metrics potentially lead to physically relevant solutions if the inhomogeneous function is physically well founded. However, according to Theorem \ref{T1}, one must return to the full system of coupled equations (Eqs. \ref{Gauss_Eq}--\ref{f_constraint_Eq}) in order to use the Hasimoto form (Eq. \ref{has}).


\section{The Nonlocal and Nonlinear Schr\"odinger Equation as an Inhomogeneous Space-Time Patching Model \label{Sec_Motiv}}

In this section, we present some motivations for the application of the NNSE/DNNSE (Eqs. \ref{NNSE}, \ref{DNNSE}) in the context of {\it inhomogeneous patching models} in LQC \cite{Boj15}. An outline of this model is given in Sec. \ref{Sec_Patch_NNSE}. The main idea is to decompose an evolving slice of the Universe into microscopic, nearly homogeneous patches, which follow a Friedmann-like dynamics, resembling a many-body, weakly interacting system. 

In a previous work \cite{Dan13}, we have indicated the similarities between the discrete NSE (DNSE, Eq. \ref{DNSE}) and a loop quantum cosmology constraint equation (for vacuum Bianchi I separable models), so that methods of solution and integrability, well-known in related Heisenberg/NSE systems, would be available for similar LQC equations. In the context of a dust matter Hamiltonian (linear in the momentum variable, conjugate to a matter degree of freedom), a nonlinear differential-difference equation (DLQC) for the single-patch wave function was proposed in Ref. \cite{Boj12} as:
\begin{equation}
i \hbar {\partial q_n \over \partial t} = (q_{n+1} - 2q_n+ q_{n-1}) +   (\Delta n)^2_q  (q_{n+1} + q_{n-1} ),~~~~~~{\rm (DLQC)}\label{DLQC}
\end{equation}
\noindent where a {\it nonlocal term} (the discrete quadractic convolution, depending on all values of the wavefunction $q_n$) is defined here as:
\begin{equation}
(\Delta n)^2_q \equiv {\alpha \over n^2}\sum_m (\langle n \rangle _q - m )^2 |q_m|^2, \label{Eq_delta_n}
\end{equation}
\noindent with the constant $\alpha$ playing a role in interacting space-time patching dynamics in analogy to the potential strength in pointlike interactions between the particles in a Bose--Einstein condensate. 

Note that nonlocality (Eq. \ref{Eq_delta_n}) is necessarily present in our change of perspective, from the interaction potential arising in many-body systems in condensed matter, to the framework of Einstein's equation (describing interacting patches of a spacelike geometry). The nonlocality is an expression of the interacting geometry being a polynomial in the metric and its spatial derivatives, shown to lead to a form like Eq. \ref{Eq_delta_n} \cite{Boj12}. 

In the following subsections, we compare the DLQC (Eq. \ref{DLQC}) and the DNNSE (Eq. \ref{DNNSE}), and discuss the NNSE (Eq. \ref{NNSE}) solutions in terms of possible  ``mesoscopic'' inhomogeneous patching models, which raises the question of considering the NNSE as interesting effective model for LQC.

\subsection{The DNNSE and the DLQC: inhomogeneities and nonlocality \label{Sub_Non}}

We consider a time-evolving spatial slice composed of small homogeneous patches. In order to include inhomogeneities, several considerations, already given in Ref. \cite{Boj12}, are here assumed.  The most relevant assumptions for the present work are small amplitude inhomogeneities, so that correlations are not strong between patches. This allows one to write the full state as a product of individual states and map the many-body dynamics to one-particle wave dynamics. 

Given the nonlocality requirement in association with a nonlinear Schr\"odinger-like equation as a driving motivation for modeling inhomogeneous space-time patching dynamics, the present investigation looked for generalizations of the DNSE (Eq. \ref{DNSE}). In the case of spin systems in condensed matter theory, a simplified model considers inhomogeneities either arising from a distribution of spin coupling strenghts, or alternatively from irregular lattice configurations, leading in either case to a variation in the overlaping of electronic wavefunctions. This model is physically well-motivated in the continuous limit by the inhomogeneous Heisenberg spin chain (Eq. \ref{S_cont}), with the associated NNSE (Eq. \ref{NNSE}). See, e.g., Balakrishnan (1982) in Ref. \cite{NLS-GEN} for further physical motivations behind this model. 

In the present work, the counterpart of the atomic lattice is the space-time patching model, quantized to the discrete parameter $n$. Given this analogy, it seems reasonable to consider the discrete form of the NNSE, the DNNSE (Eq. \ref{DNNSE}) as an alternative to, or perhaps a generalization of, the DLQC (Eq. \ref{DLQC}). Both these equations implement inhomogeneities by adding a nonlocal contribution to the nonlinear term, but the mathematical implementation is different. 

In the DNNSE (Eq. \ref{DNNSE}), lattice fluctuations (or, alternatively, varying spin coupling strenghts) are encoded in the inhomogeneity function $f_n$, which is an appropriately scaled measure of the ``fluctuation''  $\delta n/\langle n \rangle$ in $n$. This function also defines a measure of nonlocality by the associated ``response function'', $\Delta f_{n,m}$. The latter is related to {\it local changes} in the inhomogeneity. For instance, if the inhomogeneity changes abruptly (e.g., as a step function) between neighbouring sites,  the response function is a Dirac's delta function, meaning that the nonlinear response is highly local. As the inhomogeneity transitions are smoother, the response function becomes broader: the nonlinearity becomes nonlocal. Evidently, nonlocality depends also on the relative scales of the response function to the wavefunction profile. For this reason, $f_n$ linearly weights the one-particle wavefunction over overlapping nearby sites. 

In the DLQC (Eq. \ref{DLQC}), on the other hand, nonlocality is represented by the scaled variance in the distances in ``atomic patch-picture'', given by Eq. \ref{Eq_delta_n}. The larger the variance, the larger the coefficient of the nonlinear term. Nonlocality in this case does not depend on how abruptly the fluctuations change from site to site, only on their relative (quadratic) amplitudes. 
 
\subsection{The DNNSE and the DLQC: qualitative correspondences \label{Sub_Compare}} 

We briefly discuss in which sense the DNNSE (Eq. \ref{DNNSE}) and the DLQC (Eq. \ref{DLQC}) may represent similar wavefunction evolutions is some appropriate limit. Assuming  small inhomogeneities (as in Ref. \cite{Boj12}), then $f_n \approx $ constant (or $\approx 1$ by a rescaling of the time variable in the DNNSE), and the linear coefficients of the DNNSE and the DLQC equations match. A further requirement is to map the nonlinear terms of both equations, namely: 

\begin{equation}
f_n | q_n | ^2  + \sum_m \Delta f_{n,m} | q_m| ^2   \longrightarrow {\alpha \over n^2}\sum_m (\langle n \rangle _q - m )^2 |q_m|^2. \label{RESP_map1}
\end{equation}

A comment is necessary at this point. The nonlinear term $(\pm 2|q_m|^2)$ associated with the Schr\"odinger equation with a cubic nonlinearity (e.g., Ablowitz, Prinari \& Trubatch, 2004 in Ref. \cite{Lak7677}, which is the type of local nonlinearity that we are considering here), is missing the proposal of the DLQC (Eq. \ref{DLQC}), given originally in Ref. \cite{Boj12}. On the other hand, in the exact limit ($f_n \mapsto 1$,  $\Delta f_{n,m} \mapsto 0$), the DNNSE becomes the DNSE (Eq. \ref{DNSE}), but this would be true for the DLQC only if the nonlinear/local term $(\pm 2|q_m|^2)$ is introduced in that equation as well. Here consider that is so, and the relevant mapping is:

\begin{equation}
\Delta f_{n,m}  \longrightarrow {\alpha \over n^2} (\langle n \rangle _q - m )^2. \label{RESP_map2}
\end{equation}

Instead of considering specific realizations of that mapping, let us qualitatively analyse two extreme cases for the response function:

\hspace{1cm} 1-- If the inhomogeneities change very smoothly from site to site (``weak coupling transition''), the response function is broad relatively to the wavefunction, but it also has almost negligible amplitude. In this case, the first term in the nonlinear coefficient of the DNNSE (inside the large parenthesis in Eq. \ref{DNNSE}) will dominate the second (nonlocal) term, and the resulting coefficient would be ``weakly nonlocal'' (``mostly local''), say:

\begin{equation}
f_n | q_n | ^2  + \sum_m \Delta f_{n,m} | q_m| ^2   \approx | q_n | ^2+ \epsilon_n , \label{RESP_case1}
\end{equation}

\noindent where $f_n \approx 1$ and the function $\epsilon_n$ is very small for all $n$. The variance (Eq. (\ref{Eq_delta_n})) representing the nonlocal contribution in the DLQC (Eq. \ref{DLQC}) would also be very small. The DNNSE and the DLQC  would then give qualitatively similar evolutions, if this appropriately tuned regime meets the mapping requirement given by Eq. \ref{RESP_map2}. 

\hspace{1cm} 2-- If the inhomogeneities, despite being of small amplitude throughout, change abruptly from site to site, in a discretized manner (``quasi-homogeneity'' with ``strong coupling transition''), then the response function will be characterized by a series of delta functions (a ``Dirac comb'') and the nonlinearity will be locally modulated by the wavefunction profile. In other words, for a given $n$ (and $f_n \approx 1$):

\begin{equation}
f_n | q_n | ^2  + \sum_m \Delta f_{n,m} | q_m| ^2 \approx | q_n | ^2  + \sum_m \epsilon \delta_{n,m} | q_m| ^2  \approx (1+\epsilon) | q_n | ^2, \label{RESP_delta}
\end{equation}

\noindent and the local nonlinearity is enforced by a scaling factor $\epsilon$. The same effect for the DLQC (Eq. \ref{DLQC}) does not seem obvious, as this equation considers inhomogeneities, {\it but not how they locally connect}.

The above considerations point to the possibility that the DNNSE (Eq. \ref{DNNSE}) encodes in a more general way the effects of nonlocality through the  response function, with the appropriate scaling given by the inhomogeneity function. Although we did not give a rigorous proof that both equations (Eqs. \ref{DLQC} and \ref{DNNSE}) correspond to similar evolutions, in appropriate conditions, our a qualitative evaluation seems reasonable. The precise validity of those assumptions would require a numerical analysis, which goes beyond the present scope of the paper. 

\subsection{Interpretation of the NNSE solutions as ``mesoscopic'' LQC solutions \label{Sec_Interp}}

In the change of perspective from condensed matter to quantum cosmology, the {\it continuous} NNSE (Eq. \ref{NNSE}) would be equivalent to a ``mesoscopic'' space-time patching model. The main idea is to consider the NNSE as an {\it effective} description of inhomogeneous LQC, supposedly arising from unknown microscopic
quantum degrees of freedom in a full LQG theory. Any applications of the NNSE in such a context must be compatible with consistency conditions imposed by covariance and quantum state properties arising in quantum cosmology \cite{Boj06}, \cite{Boj09}, \cite{Boj12b}, so these will be discussed in the next subsection.

 In the NNSE, the inhomogeneity function must be chosen amongst those which lead to solutions (Theorems \ref{T1} and \ref{T2}), and at the same time it must also be motivated by the physical application. In this sense, models based on the NNSE may represent an interesting effective approach to LQC, because the resulting wavefunction solutions are consistently dependent on the choice of inhomogeneity. In other words, once the inhomogeneity is defined, a semiclassical quantum state is selected in this interpretation.

Evidently, the existence of exact solutions (either soliton or geometric-based, c.f. Eqs. \ref{Soliton_Sol}--\ref{Soliton_Q}; Theorems \ref{T1}, \ref{T2}) is a technical advantage in the sense that they can be treated and manipulated immediately in the theory, and be compared, e.g., with independent numerical investigations (e.g., Ref. \cite{NUMER}). The availability of a well-defined way in which the NNSE state evolves (exact solutions), even in the more complicated scenario of a time-dependent inhomogeneity function (c.f. Theorem \ref{T1}), offers an interesting opportunity for investigation. Such a description may hold at high curvature regimes, e.g., under an adequately modeled nonlocality (expressed in terms of the inhomogeneity response function, see Sec. \ref{Sub_Non}). 

For soliton solutions, one may interpret that dynamical trajectories follow
mean values in the given semiclassical quantum state. A systematic study of such evolutions may provide a qualitative basis for indicating expectation values of physical observables which may deviate from classical solutions. 
Another important issue is the fact that effective equations in general offer an approximation to a full quantum dynamics for highly peaked states. The NNSE-based models associated with soliton solutions (Eqs. \ref{Soliton_Sol}--\ref{Soliton_Q}) may offer consistent alternatives to peaked states other than Gaussians, where there is no clear underlying justification in the quantum cosmology setting \cite{Boj12b}.

As any mathematical equation, parameters of NNSE solutions can be changed arbitrarily, and by themselves do not give any predictions: they are tuned according to the application. Therefore in specific applications, the NNSE variables $\{ x, t \}$ should be interpreted accordingly in the space of expected values. For example, position $x$ may represent the physical volume ${V}$ of a ``microscopic'' cell of a classical spatial manifold, and some physical variable may play the role of an ``emergent physical time'', say, the cosmological scalar factor ${a}$. The corresponding interpretation, however, is not simple \cite{Boj12b}. The NNSE solution in this case is assumed to represent a semiclassical state, $\Psi({V},{a})$,  of a consistent and gauge invariant quantum cosmology model. That is,  $\Psi({V},{a})$ would represent  an approximation to a state that can be written in terms of the eigenfunctions of the dynamical evolution operator, characterizing the corresponding discrete model (in this case, the DNNSE, Eq. \ref{DNNSE}). However, these eigenfunctions need not to be known for constructing our space-time patching model, as the NNSE solutions are readily interpreted as representing a semiclassical quantum cosmological state. This is an assumption of the modeling, which nevertheless may lead to a starting point for a phenomenological framework to quantum cosmology. 

We briefly outline two main complementary approaches for the application of the NNSE in the context of inhomogeneous LQC (to be qualitatively explored in Sec. \ref{Sec_GETool}): 

\begin{enumerate}
\item{{\it ``Bouncing toy models''.} This route can be developed by investigating which families of solutions of the NNSE share common behaviors associated with independently known, or expected, features of LQC models, representing quasiclassical states. For instance, a general expectation of LQC is the replacement of the Big Bang singularity with a quantum ``bounce'' \cite{BOUNCE}, \cite{Cra13}.  Certain physical predictions are of great interest, e.g.,  the resulting bounds to maximum (critical) density of matter at ``bouncing''.  Considering small inhomogeneities, NNSE solutions may be initially selected based on qualitative compatibility with those scenarios. As such, they could serve as start-up ``toy models'' of quantum space-time patchings, tuned to evolve into the high curvature/density regime, exploring the impact of various inhomogeneity functions on the quantum cosmology dynamics.}
\item{{\it Consistency conditions.} This investigation concerns the possible connection of the theory of integrability of the NNSE itself, with specific constraints (consistency conditions) arising in effective LQC \cite{Boj09}, \cite{Boj12b}, under the assumption that the space-time patchings dynamics behaves as an integrable many-body system. }
\end{enumerate}

Finally, we would like to briefly comment on another, quite different, interpretation of our framework, in a more generic situation  than the dust matter model here considered. We refer to the well-known  {\it ``problem of time''} \cite{Boj11} \cite{Boj04}, arising in the Dirac quantization scheme applied to the classical Hamiltonian constraint of general relativity. In the usual LQC procedure of quantization (via symmetry reduction techniques \cite{LQC}), one adopts  internal clocks  envisaged as measures of a relational time. Now considering the absence of a matter degree of freedom (which usually plays the role of an internal time), and still considering the validity of the NNSE as an effective quantum cosmological description of that case, we reframe our interpretation via time-independent NNSE solutions (c.f. Theorem \ref{T2}). These solutions are represented by surfaces of revolution, ``internally'' parametrized by a ``time'' variable (c.f. Eq. \ref{surfrev_eq}). In this case, the function $\Phi$, related to the inhomogeneous degrees of freedom, plays the role of the generator of revolution. Therefore, this seems to raise an interesting prospect for describing inhomogeneous vacuum solutions in our effective description. A reformulation of the ``problem of time'' in such a context, however, requires a deeper, separate analysis, which is  beyond the scope of the present work.

\subsection{The LQC patchwork space-time model and the NNSE \label{Sec_Patch_NNSE}}

The usual  space-time patchwork of LQC is described in terms of a time evolving spatial slice, composed of $\mathcal{N}$ ``microscopic'' patches\footnote{We quote Bojowald in Ref. \cite{Boj09}: a patch is  ``the smallest building block of a discrete geometry''.}  with a cosmological classical geometry. Following Ref. \cite{Boj12} (see also Ref. \cite{Boj06} for further details), an inhomogeneous and isotropic (approximatelly flat) classical spatial metric, $h_{ab}$ (for scalar modes in longitudinal gauge), is given by:
\begin{equation}
h_{ab} = a(\tau)^2 \delta_{ab} + 2 ~ \mathcal{F}(x,y,z,\tau) \delta_{ab},  \label{Eq_metric}
\end{equation}
\noindent where $a$ is the cosmological scale factor, in terms of the proper time, $\tau$. The {\it metric inhomogeneity function}, $\mathcal{F}(x,y,z,\tau)$,  is discretized into a function, $\Delta \mathcal{F}$, expressing deviations of the {\it geometrical (physical) volume of a patch}, $V_{i,j,k} \in \mathcal{N}$ from the {\it total geometrical (physical)  volume} $\tilde{V}$ of the slice, as $V_{i,j,k}$ varies from patch to patch in a discrete manner, at some fixed proper time\footnote{Indices label patches in each spatial direction; associated quantities may be defined, e.g., at a central point in the patch region.}.  

An approximated expression for $\Delta \mathcal{F}$ was derived in Ref. \cite{Boj12} (c.f. their Eq. (2)). By choosing a lapse function $N = 1- 2 \mathcal{F}/a^2$, and considering nearly isotropic patches, with sizes smaller than the spatial inhomogeneities, they found:
\begin{equation}
\Delta \mathcal{F} \approx \mathcal{F}_{\rm pat}(\mathcal{V}_{\rm pat},a) = {\mathcal{V}_{\rm pat}- \tilde{V}/\mathcal{N} \over 3  a\ell_{0}^3}, \label{class_ihn_disc}
\end{equation}
where $\ell_{0}^3$ is the {\it coordinate (comoving) volume} of the patch, and $\mathcal{F}_{\rm pat}$ approximates  the integration of the function $\mathcal{F}(x,y,z,\tau)$ over the patch by its value at some point, $x_{(i,j,k)}$, of the patch (e.g., its center).
We use a simplified notation for the indices (${\rm pat} \equiv (i,j,k)$), so that $V_{i,j,k} \equiv \mathcal{V}_{\rm pat}$ and $\mathcal{F}[x_{(i,j,k)}] \equiv \mathcal{F}_{\rm pat}$. The total volume of the spatial slice is $\tilde{V} = a^3\ell_{0}^3\mathcal{N}$. 

First, we consider the limit of short evolution proper times $\tau$ of the wavefunction, as compared to the cosmological scale factor, that is, $a \approx$ constant.  In order that the NNSE (Eq. \ref{NNSE}) be an effective description of inhomogeneous LQG, we make the association of the NNSE inhomogeneity function, $f(x)$ (Eq. \ref{f_eq}), with the metric inhomogeneity function, $\mathcal{F}_{\rm pat}$ (Eq. \ref{class_ihn_disc}), at a ``mesoscopic'' limit, and consider $\mathcal{F}_{\rm pat}(\mathcal{V}_{\rm pat},a) \approx \mathcal{F}_{\rm pat}(\mathcal{V}_{\rm pat})$. In other words, if the inhomogeneity is sufficiently small and the patches smoothly connect to each other (e.g., the inhomogeneity wavelength modes are much larger than the Planck length, and produce smooth ``response functions''; see Secs. \ref{Sub_Non}, \ref{Sub_Compare}), in a weakly correlated regime, and the in the regime of short evolution proper times of the semiclassical wavefunction, then the  ``mesoscopic'' correspondence,
\begin{equation}
\mathcal{F}(x,y,z,\tau) \rightarrow \mathcal{F}_{\rm pat} ~~~~\longleftrightarrow ~~~~ 
f(x) \rightarrow f_n  ,
\end{equation}
\noindent is assumed valid, up to some factor. A corresponding association with the ``response function'' would be (c.f. Eq. (10) in Ref. \cite{Boj12}):
\begin{equation}
{\partial \over \partial x^b} {\mathcal{F} \over a^2} \rightarrow 
{\mathcal{V}_{\rm pat+\tilde{b}} - \mathcal{V}_{\rm pat-\tilde{b}} \over 6 \ell_0(\tilde{V}/\mathcal{N})}
~~~~\longleftrightarrow ~~~~
{\partial f \over \partial x}  \rightarrow \Delta f_{n,m},
\end{equation}
\noindent where $\tilde b$ is the unit vector in the $b$ direction. Considering our interpretation of the NNSE solutions as ``mesoscopic'' LQC solutions (c.f. Sec. \ref{Sec_Interp}), we map the $x$ variable in the NNSE parametrization space to the ``mesoscopic'' physical volume. For concreteness, we assume that a lower bound to this ``mesoscopic'' scale is the patch volume itself, so that we make the identification:
\begin{equation}
\mathcal{V}_{\rm pat}  ~~~~\longleftrightarrow~~~~ x \rightarrow n ~~~~~~~{\rm (mesoscopic~limit)}  \label{Eq_meso}
\end{equation}

Then, in the context of solutions to the NNSE via surface theory, according to Theorem \ref{T2}, a time-independent solution to the NNSE exists for any inhomogeneity function satisfying Eq. \ref{f_eq}, with the function $\Phi \equiv \Phi(\mathcal{V}_{\rm pat})$ representing a {\it semiclassical model for the distribution of physical  patch volumes}. In other words, our model naturally implements an inhomogeneous distribution of elementary (discrete) volumes. The corresponding solution to the NNSE is interpreted as representing the state $\Psi(\mathcal{V}_{\rm pat})$ (c.f. Sec. \ref{Sec_Interp}), that is, a semiclassical state, obtained in a self-consistent way, corresponding to a collection of fundamental quantum states of space \cite{LQG}. Therefore, it encapsulates  the overlaping of wavefunctions in a weakly-coupled, inhomogeneous, many-body patchwork system.

In more complicated scenarios, the scale factor $a$ may vary significantly throughout the wavefuntion evolution. Also, Eq. \ref{class_ihn_disc} could no longer hold. Yet, Theorem \ref{T1} still guarantees the existence of solutions for surface parameters obeying the Gauss-Mainardi-Codazzi equations (c.f. Eqs. \ref{Gauss_Eq}--\ref{MC2_Eq}), as long as the inhomogeneity function for the patches, $\mathcal{F}_{\rm pat}$, is assumed to be of the form given by Eq. \ref{f_constraint_Eq}, with the time varying ``mesoscopic'' function $f(x,t)$ as its counterpart. It is clear, therefore,  that in scenarios where the approximations leading to Eq. \ref{class_ihn_disc} no longer hold, it is still a requirement to parametrise the inhomogenous function accordingly, for instance, as some function $\mathcal{F}_{\rm pat}(\mathcal{V}_{\rm pat}(\tau),a(\tau))$. 
The corresponding solution (Eq. \ref{has}) to the NNSE is, therefore, interpreted as representing the semiclassical state $\Psi(\mathcal{V}_{\rm pat},a)$.

\section{A note on the integrability of the NNSE and consistency conditions in inhomogeneous LQC \label{Sec_GETool}}

In this section, we outline a connection between the integrability of the NNSE (Eq. \ref{NNSE}) and certain consistency conditions that arise in LQC, such as holonomy and inverse volume corrections (c.f. Ref. \cite{Boj09}; refer to the assumptions developed in this reference for further details). 

Large quantum geometry effects would be in conflict with the large-scale (classical) universe: consistency conditions give bounds for ruling out certain parameter choices in LQC. Our inhomogeneous model suggests that some qualitative bounds on quantum geometry effects (namely, leading to important corrections at the classical level) could be delineated from the geometric construction underlying the integrability of the NNSE, under the assumption that the patchwork dynamics behaves as an integrable many-body system.

In Ref. \cite{Boj09}, several issues have been raised on quantum geometry effects resulting from changes in the assumptions regarding the nature of the discrete geometry, how it evolves, anisotropy considerations, etc. In order to keep our arguments concrete, we present, for illustrative purposes,  the simplest consistency conditions arising from LQC, under a mean field approach. We invert Eq. \ref{class_ihn_disc} in order to obtain the physical patch volume in terms of the inhomogeneity of the patchwork:
\begin{equation}
\mathcal{V}_{\rm pat} = \tilde{V}/\mathcal{N} + 3  a\ell_{0}^3 \mathcal{F}_{\rm pat}, \label{Eq_V_pat}
\end{equation}
\noindent so that the corresponding {\it geometrical (physical) patch size length} $\mathcal{L}_{\rm pat}$ is given by:
\begin{equation}
\mathcal{L}_{\rm pat} = (\mathcal{V}_{\rm pat})^{1 \over 3} = a\ell_{0} +  (3  a \mathcal{F}_{\rm pat})^{1 \over 3} \ell_{0} = \left [ a +  (3  a \mathcal{F}_{\rm pat})^{1 \over 3}\right ] \ell_{0}. \label{Eq_L_pat}
\end{equation}

Following Ref. \cite{Boj09},  for quantum {\it holonomy corrections} to be small in the classical regime, the following condition must hold:
\begin{equation}
\gamma {\dot{a} \over a} \mathcal {L}_{\rm pat} \ll 1, \label{hol}
\end{equation}
\noindent where the dot refers to the derivative with respect to a proper time; $\gamma$ is the Barbero-Immirzi parameter \cite{IRM}. For {\it inverse volume corrections} to be small in classical regimes, we have:
\begin{equation}
{\mathcal {L}_{\rm pat} \over \ell_{\rm P}} \gg 1, \label{invvol}
\end{equation}
\noindent where $\ell_{\rm P}$ is the Planck length. On the other hand, considering effective theories in the homogeneous and isotropic sector of LQC, as well as in pertubations around these backgrounds (e.g. Ref. \cite{Wil12}), modified Friedmann equations can be derived, which reduce to the classical equations, in the limit when the {\it critical density}, 
\begin{equation}
\rho_{\rm{crit}} \equiv {3 \over 8 \pi G \gamma^2 \mathcal{L}_{\rm pat}^2}, \label{rho-crit}
\end{equation}
\noindent tends to infinity. In other words, $\rho_{\rm{crit}}$ must be large compared to actual matter densities in the early universe. A bouncing effect at $\rho = \rho_{\rm{crit}}$ (when the scale factor reaches a minimum) is a general prediction of those models \cite{BOUNCE}.  Eqs. \ref{hol} and \ref{invvol}, therefore, compose mutual conditions that quantum corrections must satisfy in order be consistent with the classical universe.

Now, the patch structure implemented in the NNSE context (Sec. \ref{Sec_Patch_NNSE}) is not aleatory. It must be connected with the existence of solutions to the NNSE, under the conditions imposed by the Theorems \ref{T1} and \ref{T2}. In other words,  the inhomogeneity of the patchwork is necessarily written in terms of an {\it admissible} $\Phi \equiv \Phi(\mathcal{V}_{\rm pat})$, or more generally in terms of admissible surface parameters, $G, L, M, N$ (c.f. App. \ref{App_surface_theory}), which will evolve in our interpretation as functions of  $\{\mathcal{V}_{\rm pat},a\}$ (i.e., they should obey the Gauss-Mainardi-Codazzi equations, c.f. Eqs. \ref{Gauss_Eq}--\ref{MC2_Eq}). 

Certain patchwork inhomogeneities will not lead to NNSE solutions. One may consider that such inhomogeneities, in a fundamental level, do not lead to a valid   ``mesoscopic'' limit, at which the NNSE admit solutions. So there is a sense in which one is able to explore various refinements and patch volume distributions, under a variety of combinations of discretized volume spectra, and test those conditions leading or not to solutions to the NNSE, in this ``mesoscopic'' sense. It would be also interesting to perform a thorough analysis of the conditions which these mesoscopic solutions arise (or not) from the continuum limit, or adopted approximations,  of the fully discretized NNSE (DNNSE, Eq. \ref{DNNSE}). For instance, such an analysis could shed light in 
certain regularization ambiguities in canonical quantum gravity \cite{Boj09}.

Such studies could be performed purely in terms of the geometrical equivalence interpretaton of the integrability of the NNSE.
In a sense, they would be analogous to the ``lattice refinement parametrization''  framework in effective theories (e.g. Ref. \cite{Boj06}), where, for instance, a model for the a time-evolving number of patches is considered and can be tuned to match phenomenology. However, in our case, the refinement is carried out in the geometrical space of the surface solution of the NNSE.
Small distortions of the function,
$\delta\Phi(\mathcal{V}_{\rm pat})$, or corresponding distortions of surface parameters, away from the exact solution surface of the NNSE, could indicate the corresponding departures at which the patchwork inhomogeneities no longer  behave as an integrable many-body system, in the adopted ``mesoscopic'' limit.  Such departures could also be investigated as a function of ``coarse-graining'' of the solution surfaces. In particular, those various types of changes are expected to have an impact on consistency conditions (Eqs. \ref{hol} and \ref{invvol}), as they affect the $\mathcal{L}_{\rm pat}$ under specific assumptions or combinations of discretized volume spectra.

For instance, in the ``bouncing'' scenario, captured by Eq. \ref{rho-crit} (with $\rho_{\rm{crit}} \rightarrow \infty$ ), an important issue is the determination of the value of the critical density in terms of non-arbitrary parameters \cite{Boj09}, that is, by some fixing (or fundamental derivation) of $\mathcal{L}_{\rm pat}$. In our model, a given inhomogeneity distribution as a function of $\mathcal{L}_{\rm pat}$ must necessarily be admissible in order to lead to a solution to the NNSE. An admissible, highly localized inhomogeneity function, for example,  leading to a soliton solution (e.g., Eqs. \ref{Soliton_Sol}--\ref{Soliton_Q}), is naturally  $\mathcal{L}_{\rm pat}$--bounded, that is,  it will correspond in the space of patch volumes to a localized geometric inhomogeneity in a small domain. 
A systematic analysis on such bounds and the impact on the consistency conditions (Eqs. \ref{hol} and \ref{invvol}) could present valuable clues on constructing admissible, many-body space-time patchwork dynamics, under the integrable hypothesis.

\section{Conclusion \label{Sec_Conc}}

In this paper, we presented a class of nonlocal and nonlinear, inhomogeneous Schr\"odinger equation (NNSE) as a space-time patching model for inhomogeneous loop quantum cosmology. We presented known classes of exact solutions of the NNSE in terms of solitons and geometric equivalence (GE) methods, and offered an interpretation of these solutions in terms of semiclassical quantum geometry wavefunctions.

We also argued that the integrability of the NNSE offers an interesting, self-consistent and relatively simple formalism for defining viable, inhomogeneous space-time patchings models. Furthemore, we outlined how the integrability of the NNSE could be linked to consistency conditions arising in LQC. We leave for a future work a more quantitative, systematic study of solution surfaces parameterized by the patch size volume and the scale factor, and how to relate them with the mutual consistency conditions imposed by quantum gravity corrections. Such a study would indicate models suitable for the microscopic space-time dynamics, which might provide interesting constraints for potential observables in quantum cosmology (e.g., Ref. \cite{SIG}).

As mentioned in Ref. \cite{Boj09}, the evolution of patch numbers and patch volumes in various phases of the universe can only be ``derived from a full solution of an inhomogeneous state, which is difficult''. The NNSE model and its integrability properties seem to offer an alternative route to this complicated problem.

\begin{acknowledgement}
We thank the anonymous referee for important corrections and useful comments, which significantly improved the contents and presentation of this work.
\end{acknowledgement}


\appendix

\section{Geometric Equivalence methods \label{App}} 

\subsection{``Moving curves'' formalism \label{App_SCF}}

\begin{center}
\begin{figure}[htbp]
\centering
\includegraphics[width=0.6\columnwidth]{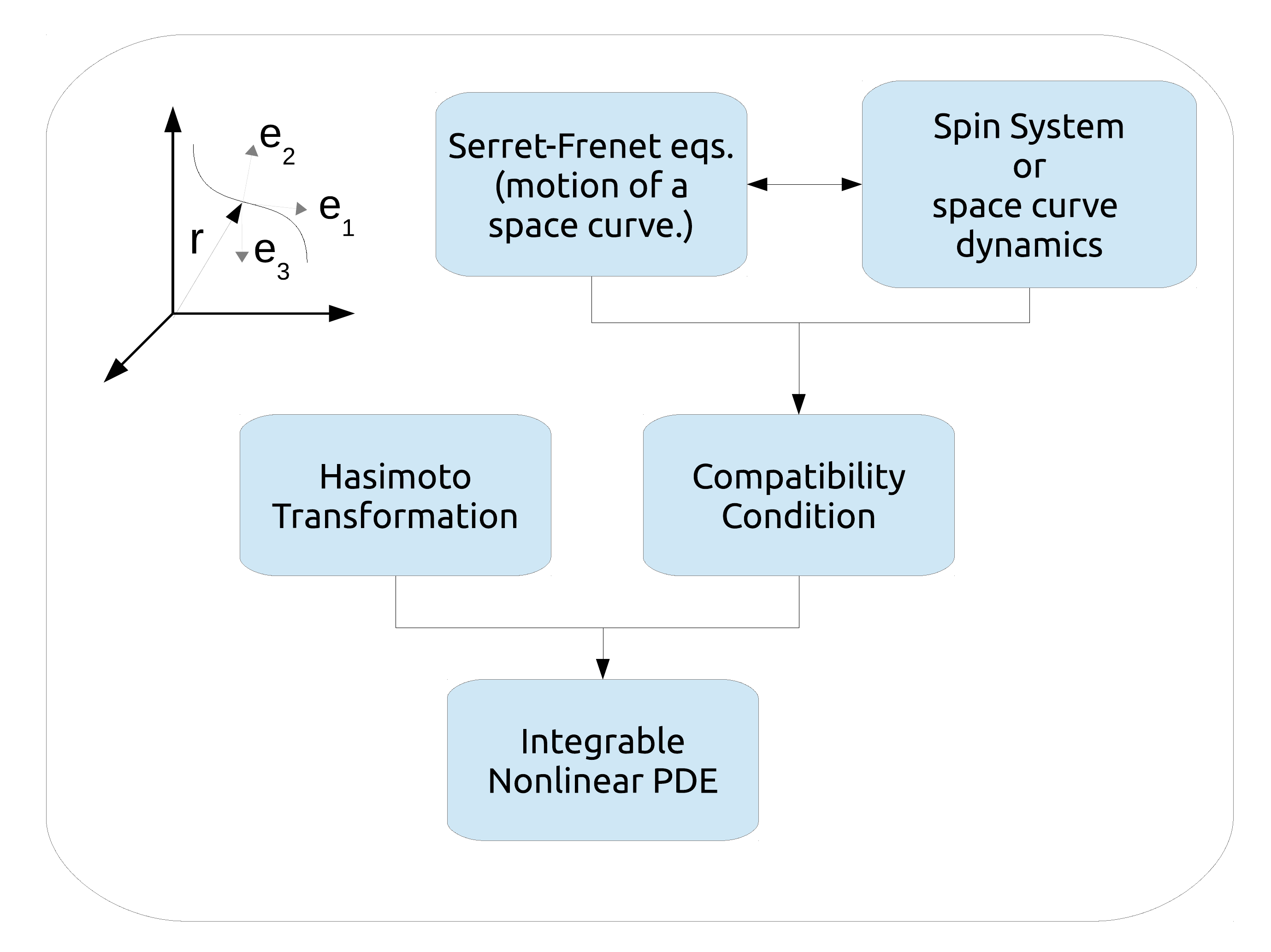}
\caption{\label{fig1} Main elements of the geometric equivalence method and their relations.}
\end{figure}
\end{center}

Our exposition here follows the general presentations given in the references \cite{GEO}, \cite{GE-REFS}. We will apply this technique to show the equivalence between a simple Heisenberg spin system and the nonlinear Schr\"odinger equation (see proofs also in, e.g., Refs. \cite{Lak11}, \cite{Lak7677}). The general scheme of the formalism can be summarized in Fig. \ref{fig1}.

Consider the 1+1-dimensional motion of a rigid spatial curve, generated by the position vector $\vec{r}(x,t) = r_1 (x,t) \vec{\hat{u}}_1 + r_2 (x,t) \vec{\hat{u}}_2 +r_3 (x,t) \vec{\hat{u}}_3$, with $x, t$ parameters in $\mathbb{R}$. The motion satisfies the Serret-Frenet equations\footnote{We identify the parameter $x$ with the arc length of the curve.}:

\begin{equation}
{\partial \over \partial x}
\left( 
\begin{array}{c} 
\vec{e}_1 \\
\vec{e}_2 \\
\vec{e}_3
\end{array} 
\right)
\mbox{~=~} 
C
\left( 
\begin{array}{c} 
\vec{e}_1 \\
\vec{e}_2 \\
\vec{e}_3
\end{array} 
\right) ~~~~~ ; ~~~~~
{\partial \over \partial t}
\left( 
\begin{array}{c} 
\vec{e}_1 \\
\vec{e}_2 \\
\vec{e}_3
\end{array} 
\right)
\mbox{~=~} 
W
\left( 
\begin{array}{c} 
\vec{e}_1 \\
\vec{e}_2 \\
\vec{e}_3
\end{array} 
\right), \label{FS_eq}
\end{equation}

\noindent with

\begin{equation}
C = 
\left( 
\begin{array}{ccc} 
0        & \kappa & 0     \\
-\kappa  & 0      & \tau \\
0        & -\tau  & 0
\end{array} 
\right)
\mbox{~~~~~; ~~~~~} 
W = 
\left( 
\begin{array}{ccc} 
0          & \omega_3   & -\omega_2     \\
-\omega_3  & 0          & \omega_1      \\
\omega_2   & -\omega_1  & 0
\end{array} 
\right), \\[15pt] \label{CW}
\end{equation}
\noindent where
\begin{equation}
\vec{e}_1 \equiv \partial \vec{r} / \partial x \label{e1_r}
\end{equation}
\noindent is the unit tangent vector at a point on the curve;  $\vec{e}_2$ is the unit normal vector, orthogonal to $\vec{e}_1$; $\vec{e}_3 \equiv \vec{e}_1 \times \vec{e}_2$ is the unit binormal vector; $\kappa$ is the curvature:
\begin{equation}
\kappa \equiv | \partial\vec{e}_1/\partial x |, \label{curvature}
\end{equation}
\noindent and $\tau$ is the torsion:
\begin{equation}
\tau \equiv (1/\kappa^2) \vec{e}_1 \cdot (\partial\vec{e}_1/\partial x \times \partial^2\vec{e}_1/\partial x^2). \label{torsion}
\end{equation}

The space curve motion can be made completely equivalent to a spin system dynamics.
Consider the simplest spin chain Heisenberg ferromagnet, described by the following equation of motion in the continuous limit (e.g., Ref. \cite{Lak11}):

\begin{equation}
{\partial \vec{S}\over \partial t} = \vec{S} \times {\partial^2 \vec{S}\over \partial x^2},  \label{Heisen} 
\end{equation}
\noindent with components satisfying $\vec{S}(x,t)= S_1(x,t) \hat{\vec{u}}_1 + S_2(x,t)\hat{\vec{u}}_2 + S_3(x,t)\hat{\vec{u}}_3$, $\vec{S}^2(x,t)= 1 ~\forall x,t$, and $\hat{\vec{u}}_i$ a normalized basis in the spin vector space. Identifying:
\begin{equation}
\vec{e}_1 \equiv \vec{S} \label{e1_S}
\end{equation}
\noindent and inserting Eqs. \ref{FS_eq} and \ref{CW} into Eq. \ref{Heisen}, one can show that the $\omega$ parameters must be related to the curvature and torsion by:

\begin{equation}
\left\{ 
\begin{array}{rcl}
\omega_1 & = & {\displaystyle-{1 \over \kappa}{\partial^2 \kappa \over \partial x^2} + \tau^2}, \\[5pt] 
\omega_2 & = & {\displaystyle{\partial \kappa \over \partial x}},\\[5pt]
\omega_3 & = & \kappa \tau .
\end{array}
\right. \\[15pt] \label{spin_constr}
\end{equation}

There is one further condition that the identification above must obey, namely: the compatibility condition between the partial derivatives of the Serret-Frenet equations. The latter imposes relationships for the parameters of the matrices $C$ and $W$:

\begin{equation}
{\partial C \over \partial t} - {\partial W\over \partial x} + [C,W] = 0
~\Rightarrow ~ 
\left\{ 
\begin{array}{rcl}
{\displaystyle {\partial \kappa \over \partial t}}  & = & {\displaystyle {\partial \omega_3 \over \partial x} + \tau \omega_2}, \\[5pt] 
{\displaystyle {\partial \tau \over \partial t}}    & = & {\displaystyle {\partial \omega_1 \over \partial x} - \kappa \omega_2 },\\[5pt]
{\displaystyle {\partial \omega_2 \over \partial x}} & = & {\displaystyle \tau \omega_3 - \kappa \omega_1}.
\end{array}
\right. \\[15pt]
\label{compat}
\end{equation}

By inserting the spin-curve identification (Eq. \ref{spin_constr}) into the compatibility condition equations (Eq. \ref{compat}), one eliminates the $\omega$ parameters and arrives at the following coupled equations involving $\kappa$ and $\tau$ only:

\begin{equation}
\left\{ 
\begin{array}{rcl}
{\displaystyle {\partial \kappa \over \partial t} - {\partial(\kappa \tau)\over \partial x} -{\partial \kappa \over \partial x} \tau} & = & {\displaystyle 0}, \\[10pt]
{\displaystyle {\partial \tau \over \partial t} + 
{\partial \over \partial x}  \left ( {1 \over \kappa}  {\partial^2\kappa \over \partial x^2} \right ) -2 \tau {\partial \tau \over \partial x} + \kappa {\partial \kappa \over \partial x}} & = & {\displaystyle 0} .
\end{array}
\right. \\[15pt] \label{coupled}
\end{equation} 

The curvature and torsion parameters satisfying the coupled differential equations above correspond to an exactly solvable and integrable rigid curve motion for the present application (the spin system given by Eq. \ref{Heisen}), and it turns out to be true for several other applications as well. The final step then consists of using a transformation which will relate the former parameters to a soliton solution of an integrable PDE counterpart of the spin system. One can show that the following so-called {\it Hasimoto transformation} will do the job:
\begin{equation}
q (x,t)  = {\kappa \over 2} \exp \left ( -i \int \tau dx \right ) . \label{hasimoto_transf}
\end{equation}
\noindent It can be verified, with Eq. \ref{coupled} and by direct substitution of the Hasimoto transformation into the nonlinear Schr\"odinger equation (NSE, Eq. \ref{NSE}), that indeed this transformation corresponds to the wave function solution $q(x,t)$ of the NSE \footnote{Other possibilities instead of the Hasimoto transformation were found and discussed in, e.g.,  Murugesh and Balakrishnan (2001) \cite{GE-REFS}. In the literature, the identification given by Eq. \ref{spin_constr} is sometimes written with opposite sign, leading to a positive sign in the exponential function of the Hasimoto transformation.}.

The Heisenberg spin chain described by the continuous spin equation of motion (Eq. \ref{Heisen}) is valid when the spin and coupling vary slowly over one lattice separation; the associated discrete equation of motion is:

\begin{equation}
{\partial \vec{S}_n\over \partial t} = \vec{S}_n \times \vec{S}_{n+1} + \vec{S}_n \times \vec{S}_{n-1}, ~~~,~~~\vec{S}^2 = 1.  \label{Heisen_eqs_motion_disc} 
\end{equation}
\noindent where $\vec{S}_n$ is the classical spin vector at site $n$. 

A proof of the equivalence of the spin system above (Eq. \ref{Heisen_eqs_motion_disc}) and the {\it discrete} nonlinear Schr\"odinger equation (DNSE, Eq. \ref{DNSE}) can be found in Hoffmann (2000, in Ref. \cite{DISC}). Hoffmann uses a different technique than the {\it discrete} ``moving curves'' method, developed by Doliwa \& Santini (1995, in Ref. \cite{DISC}). The literature on the discrete ``moving curves'' method is much more limited than the continuous case. To the author's best knowledge, there is no straightforward proof of the equivalence between Eqs. \ref{Heisen_eqs_motion_disc} and \ref{DNSE} in the literature by the use of {\it discrete GE} methods. As mentioned, Doliwa \& Santini did establish the conditions where the motion of a piecewise linear curve in $\mathbb{R}^3$ leads to integrable discrete equations. In this way, they have illustrated the correspondence between discretized versions of the NSE and the motion of discrete curves. See their Eq. (89) and a brief mentioning of its correspondence with an analogous discrete spin system (the Heisenberg XXO antiferromagnet). In few subsequent works by Daniel \&  Manivannan (1998, 1999, in Ref. \cite{DISC}), using the results by Doliwa \& Santini, the integral motions of discretized curves were directly associated with spin equations of motion. These authors describe the procedure for closely related systems to that of Eq. \ref{Heisen_eqs_motion_disc}: the Ishimori spin systems and higher order spin systems. 

A generalization of the continuous spin system (Eq. \ref{Heisen}) introduces {\it inhomogeneity} or {\it coupling}, described by a time-independent function $f(x)$:
\begin{equation}
{\partial \vec{S} \over \partial t} = {\partial \over \partial x} \left [ f(x) \left ( \vec{S} \times {\partial \vec{S} \over \partial x} \right ) \right ].~~~~~~\vec{S}^2 = 1. \label{S_cont}
\end{equation}
\noindent It can be shown that the latter continuous spin system corresponds to the nonlocal (inhomogeneous) nonlinear Schr\"odinger equation (NNSE, Eq. \ref{NNSE}; c.f. Balakrishnan (1982, 1982b) in Ref. \cite{NLS-GEN}). 

A related {\it discrete and inhomogeneous} spin system can be written as:
\begin{equation}
{\partial \vec{S}_n \over \partial t}  = f_n (\vec{S}_n \times \vec{S}_{n+1}) + f_{n-1}  (\vec{S}_n \times \vec{S}_{n-1})~~~,~~~\vec{S}_n^2 = 1, \label{S_disc}
\end{equation}
\noindent where $f_n$ corresponds to a site-dependent nearest-neighbor interaction function in the discrete model. To the author's best knowledge, the correspondence between the discrete inhomogeneous spin equation above and the discrete version of the NNSE (namely, the DNNSE, Eq. \ref{DNNSE}) has not been studied. Such a correspondence, however, can in principle be established by the methods of Doliwa \& Santini (1995, in Ref. \cite{DISC}) and Daniel \&  Manivannan (1998, 1999, in Ref. \cite{DISC}).

\subsection{Surface theory \label{App_surface_theory}}

Geometric equivalence methods can also be developed in terms of {\it surface theory} instead of ``moving curves''. The vector $\vec{r}(x,t)$ defining a moving curve can also be regarded as the generator of a surface in $\mathbb{E}^3$. Such a surface is not arbitrary, as it is produced by a special (nonstretching) kind of curve.

Here we briefly state the basic results. The metric on the surface is given by
\begin{equation}
(d\vec{r})^2 = E (d x)^2 + 2 F dx dt + G (dt)^ 2,
\end{equation}
\noindent with $E = (\partial \vec{r}/ \partial x)^2$, $F = \partial \vec{r}/ \partial x \cdot \partial \vec{r}/ \partial t$, and $G =( \partial \vec{r}/ \partial t)^2$. Defining the unit normal vector, now orthogonal to the surface, as 
$
\vec{n} \equiv (\partial \vec{r}/ \partial x \times \partial \vec{r}/ \partial t) / |\partial \vec{r}/ \partial x \times \partial \vec{r} / \partial t |,
$
one has the moving triad of linear independent vectors $(\partial \vec{r}/ \partial x ,\partial \vec{r}/ \partial t, \vec{n})$. Furthermore, one also has the extrinsic curvature tensor, defined by:
\begin{equation}
T \equiv - d\vec{r}\cdot d\vec{n} = L (dx)^2 + 2 M dx dt + N (dt)^2.
\end{equation}

As previously done for curves, one identifies the spin vector with a geometrical object (c.f. Eq. \ref{e1_S}): ${\partial \vec{r} \over \partial x} \equiv \vec{S}$, and the model requires $\vec{S}^2 = 1$, which implies $E=1$ and $F=0$ in the present case. Then, according to the spin dynamical equation of interest, the corresponding kinematic equation relating $\vec{r}$ and its derivatives can be written down explicitly. 

For instance, Balakrishnan \& Guha \cite{Bal96} (see also Ref. \cite{Bal97})
have shown that an exact solution to the inhomogeneous spin dynamics, Eq. \ref{S_cont}, exists, provided that the metric function $G$ and the extrinsic curvature functions $L$, $M$ and $N$ satistify the following system :

\noindent\hspace{1cm} 1) the Gauss equation,
\begin{equation}
-K=(G_x/2G)_x+(G_x/2G)^2,  \label{Gauss_Eq}
\end{equation}
\noindent or its equivalent form,
\begin{equation}
(G^{1/2})_{xx}= -KG^{1/2},  \label{Gauss2_Eq}
\end{equation}
\noindent where the Gaussian curvature is defined as:
\begin{equation}
K=(LN-M^2)/G;  \label{Gauss_Curv_Eq}
\end{equation}
\hspace{1cm} 2) the Mainardi-Codazzi equations,
\begin{equation}
L_t-M_x=M(G_x/2G),  \label{MC1_Eq}
\end{equation}
\begin{equation}
M_t-N_x=-L(G_x/2)+M(G_t/2G)-N(G_x/2G); \label{MC2_Eq}
\end{equation}
\hspace{1cm} 3) the constraint for the inhomogeneous function,
\begin{equation}
f = - G^{1/2}/L. \label{f_constraint_Eq}
\end{equation}

Consider the simplest nontrivial case: an arbitrary time-independent metrics, namely $G(x,t) = G(x)$ (or $G_t=0$), and time-independent coupling, $f(x,t)=f(x)$, both implying $L_t=0$ (from Eq. \ref{f_constraint_Eq}). This, in turn, makes it possible to integrate for $M$ in $x$ (Eq. \ref{MC1_Eq}), setting the arbitrary integration constant (say, $C_0$) as independent of time (and then $M_t=0$). Finally, Eq. \ref{MC2_Eq} can also be simplified and integrated in $x$. By assuming these simplifications, one can obtain explicit solutions satisfying the above set of equations, in other words, one obtains a set of constraints on the admissible geometries of the generated surface. See discussion in Sec. \ref{Sec_GE_NNSE}. 

The relation between the moving curve parameters which generates the surface is:
\begin{equation}
\kappa = L,
\end{equation}
\noindent and
\begin{equation}
\tau = M/G^{1 \over 2}.
\end{equation}
\noindent Using those relations in the Hasimoto transformation eq. \ref{hasimoto_transf}, one obtains Eq. \ref{has}.

An intuitive example (see, e.g., Murugesh \& Lakshmanan, 2005 \cite{GE-REFS}) of how surfaces and partial differential equations can be made to correspond each other is to consider a pseudo spherical surface (Gaussian curvature of $-1$). Then, the angle $\theta$ between the surface tangent vectors $\partial \vec{r}/ \partial x$ and $\partial \vec{r}/ \partial t$ is known to satisfy: $\theta_{xt} = \sin \theta$. But this is just the well-known integrable sine-Gordon equation.


\end{document}